%% file: main.tex
\documentclass[10pt]{article}
\usepackage[utf8]{inputenc}
\usepackage[T1]{fontenc}
\usepackage{times}
\usepackage[margin=0.78in]{geometry}
\usepackage{amsmath,amssymb}
\usepackage{booktabs}
\usepackage{graphicx}
\usepackage{microtype}
\usepackage{xcolor}
\usepackage{url}
\usepackage{hyperref}
\usepackage{enumitem}
\usepackage{multirow}
\usepackage{array}
\usepackage{float}
\usepackage{xspace}
\usepackage{tikz}
\usetikzlibrary{arrows.meta,positioning,fit}

\hypersetup{colorlinks=true,allcolors=blue!55!black,hypertexnames=false}
\setlist{nosep,leftmargin=*}
\newcommand{\system}{BVR Sim\xspace}

\title{BVR Sim: An Open and High-Throughput Environment\\for Heterogeneous Air-Combat Reinforcement Learning}
\author{Haocheng Sun\\Beijing University of Posts and Telecommunications\\\texttt{sunhaocheng@bupt.edu.cn}
\and Mulai Tan\\Aviation Engineering School, Air Force Engineering University\\Xi'an 710038, China\\\texttt{kgdtml@163.com}}
\date{}

\begin{document}
\maketitle

\begin{abstract}
Beyond-visual-range (BVR) air combat is a challenging reinforcement-learning domain characterized by partial observability, long-horizon decision making, energy management, and limited weapons. We present \system, an open-source Gymnasium-style environment designed for heterogeneous air-combat reinforcement learning. BVR Sim supports multiple JSBSim aircraft models, including the F-15, F-16, F/A-18, and F-22, with configurable weapons, sensors, controllers, and opponents. A unified tactical action interface specifies desired heading, altitude, speed, and weapon release above aircraft-specific inner-loop controllers, enabling policies to operate across heterogeneous platforms. The environment provides interchangeable Python and accelerated C++ backends, entity-oriented observations, compositional rewards, scripted opponents, replay and visualization, and adapters for multi-agent learning frameworks. At a $0.4$-s decision interval, the C++ backend achieves 104 simulated seconds per wall-clock second in 1-vs-1 and remains practical through 10-vs-10 scenarios. A policy trained only on the F-16 transfers without retraining to four unseen aircraft, reaching a $45.5\%$ mean win rate with aircraft-specific controller adaptation. MAPPO and HAPPO experiments further verify end-to-end compatibility with standard multi-agent reinforcement-learning pipelines. Open-source code: \url{https://github.com/lizi-Margin/bvr_sim}.
\end{abstract}

\section{Introduction}
Air combat beyond visual range combines continuous dynamics with discrete, delayed, and often irreversible decisions. An aircraft must preserve energy, establish a favorable geometry, decide whether a radar track is actionable, allocate a finite missile inventory, support a launched weapon, and evade incoming threats. In team engagements these decisions are coupled across agents and occur under partial information. The domain therefore exposes several central challenges in multi-agent reinforcement learning (MARL): long horizons, sparse terminal outcomes, non-stationarity, credit assignment, heterogeneous capabilities, and coordination under imperfect sensing.

Open simulators have made aerial reinforcement learning substantially more accessible. BVR Gym uses JSBSim and provides a structured environment for BVR tactics~\cite{scukins2024bvrgym}. Light Aircraft Game (LAG) offers a lightweight Gym-wrapped aircraft combat suite and hierarchical training tasks~\cite{liu2022lag,cao2025lag}. B-ACE emphasizes an accessible, lightweight BVR setting~\cite{kuroswiski2024bace}, while WUKONG demonstrates reinforcement-learning-based BVR tactic generation~\cite{piao2020wukong}. These systems establish the value of an open air-combat benchmark, but leave a useful design space between fidelity, throughput, and learning ergonomics.

\system targets this design space as a \emph{complete environment}, rather than as a standalone missile model or flight-control module. Its principal design choice is to separate the frequency and responsibility of tactical learning from those of vehicle stabilization. The policy selects desired changes in heading, altitude, and speed together with a fire decision; model-specific flight controllers translate these commands to control-surface and throttle inputs. This separation gives the learning policy consistent tactical action semantics across aircraft whose low-level dynamics differ substantially. It avoids requiring every BVR experiment to first learn basic flight, while retaining six-degree-of-freedom JSBSim aircraft dynamics. The same environment also supports heterogeneous formations: aircraft type, flight-dynamics implementation, controller parameters, initial state, radar behavior, pylons, weapons, and opponent policy are configured per unit.

This paper makes four contributions:
\begin{enumerate}
    \item We introduce an open end-to-end BVR MARL environment that integrates scenario generation, dynamics, sensing, weapon employment, learning interfaces, baselines, and visualization.
    \item We provide heterogeneous per-unit aircraft, weapon, sensor, and controller configuration within one engagement, rather than treating heterogeneity only as different learned policies over identical vehicles.
    \item We expose a shared learning-facing tactical action abstraction and entity-oriented observations across heterogeneous aircraft, while model-specific inner-loop controllers handle actuation.
    \item We provide dual Python and C++ backends and characterize the environment through throughput, standard MARL integration, and cross-aircraft controller-transfer studies with reproducible artifacts.
\end{enumerate}

\section{Related Work}
\paragraph{Aircraft combat learning environments.}
BVR Gym combines JSBSim aircraft dynamics, missile simulation, behavior-tree opponents, and RL interfaces for BVR engagements~\cite{scukins2024bvrgym}. Its layered organization provides controllers and agent abstractions, and is an important direct reference for our environment. LAG covers single-aircraft control, single combat, and multi-aircraft combat, and uses a hierarchical controller in which a learned or specified high-level command is realized by a low-level policy~\cite{liu2022lag,cao2025lag}. In contrast, \system makes a conventional model-specific flight controller part of the simulator, so a tactical policy can be trained directly without a separately trained low-level policy. B-ACE uses the Godot engine to provide a lightweight open BVR environment~\cite{kuroswiski2024bace}. WUKONG studies automated BVR tactic generation using reinforcement learning, but its underlying simulator is not released as an open research environment~\cite{piao2020wukong}.

\paragraph{Multi-agent benchmarks and algorithms.}
General MARL benchmarks such as SMAC~\cite{samvelyan2019smac}, PettingZoo~\cite{terry2021pettingzoo}, and Multi-Agent MuJoCo~\cite{peng2021facmac} have helped standardize evaluation, but do not encode the coupled flight, sensing, energy, and weapon constraints of BVR combat. \system exports standard spaces and includes adapters for established MARL runners, allowing algorithms such as MAPPO~\cite{yu2022mappo}, HAPPO~\cite{kuba2022happo}, and off-policy methods to be evaluated without making the simulator depend on one training implementation.

\paragraph{Flight dynamics.}
JSBSim is an open flight-dynamics library widely used in research and engineering~\cite{berndt2004jsbsim}. Both BVR Gym and LAG build on JSBSim. \system also uses JSBSim for aircraft, but embeds it in a broader object, weapon, sensing, control, and telemetry architecture and offers a native C++ execution path to reduce Python overhead in large engagements.

\section{Environment Overview}
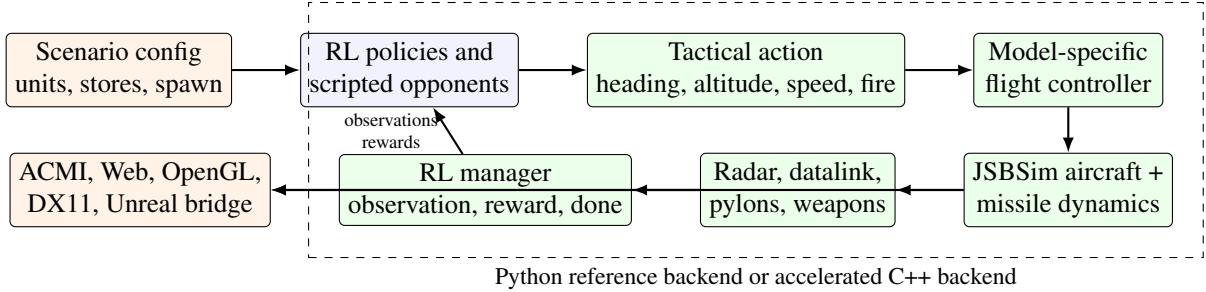
\begin{figure*}[t]
\centering
\begin{tikzpicture}[
  node distance=6mm and 9mm,
  box/.style={draw,rounded corners=2pt,align=center,minimum height=9mm,minimum width=25mm,fill=blue!5},
  core/.style={box,fill=green!8},
  io/.style={box,fill=orange!10},
  arr/.style={-{Latex[length=2mm]},thick}
]
\node[io] (config) {Scenario config\\units, stores, spawn};
\node[box,right=of config] (agents) {RL policies and\\scripted opponents};
\node[core,right=of agents] (action) {Tactical action\\heading, altitude, speed, fire};
\node[core,right=of action] (control) {Model-specific\\flight controller};
\node[core,below=of control] (physics) {JSBSim aircraft +\\missile dynamics};
\node[core,left=of physics] (systems) {Radar, datalink,\\pylons, weapons};
\node[core,left=of systems] (rl) {RL manager\\observation, reward, done};
\node[io,left=of rl] (telemetry) {ACMI, Web, OpenGL,\\DX11, Unreal bridge};
\draw[arr] (config) -- (agents);
\draw[arr] (agents) -- (action);
\draw[arr] (action) -- (control);
\draw[arr] (control) -- (physics);
\draw[arr] (physics) -- (systems);
\draw[arr] (systems) -- (rl);
\draw[arr] (rl) -- node[left,align=center,font=\scriptsize]{observations\\rewards} (agents);
\draw[arr] (physics) -- (telemetry);
\draw[arr] (systems) -- (telemetry);
\node[draw,dashed,fit=(action)(control)(physics)(systems)(rl),inner sep=4mm,label=below:{\small Python reference backend or accelerated C++ backend}] {};
\end{tikzpicture}
\caption{\system integrates heterogeneous unit configuration with a shared tactical interface and aircraft-specific inner loops. Rendering and telemetry are decoupled from the Python or accelerated C++ headless backends.}
\label{fig:architecture}
\end{figure*}

Figure~\ref{fig:architecture} shows the environment as a closed loop. A JSON/JSONC scenario instantiates red and blue aircraft, their dynamics and stores, optional ground objects, spawn geometry, and opponent types. At each environment step, policy and scripted actions are translated by the tactical action layer. Inner-loop controllers drive aircraft dynamics; sensors and weapon systems update; the RL manager constructs observations, rewards, and termination signals; and telemetry consumers optionally record or visualize the state.

The simulator advances with a configurable environment interval $\Delta t$ (the supplied benchmark scenarios use $0.4$~s). JSBSim is internally sub-stepped at no more than $0.02$~s in the C++ backend, so a tactical decision can occur at a lower frequency than numerical flight integration. This multi-rate design is important: BVR decisions do not need to be issued at actuator frequency, but aircraft stability and missile integration still require smaller steps.

\section{Heterogeneous Engagement Modeling}
\subsection{Per-unit configuration}
Each aircraft is created from a per-unit specification rather than a team-wide hard-coded model. A unit entry defines an identifier, team, three-dimensional initial position and velocity, flight-dynamics type, aircraft specification, pylon mounts, and an optional scripted opponent. The current C++ JSBSim mapping includes F-15, F-16, F/A-18, F-4N, AJ/JA-37, and F-22 families. Consequently, a single episode can combine aircraft with different aerodynamic models and controller parameters. The repository includes mixed F-15/F-16/F-22 5-vs-5 scenarios as executable examples.

Heterogeneity is not cosmetic. The aircraft model selects a JSBSim definition and a corresponding flight-controller parameter set. The unit's pylon map independently determines weapon quantity and type, including AIM-120 variants and short-range weapons supported by the weapon factory. The same mechanism permits randomized aircraft specifications at reset, enabling training over a distribution of force compositions.

\subsection{Scenario generation}
The spawn manager generates opposing formations from an initial separation, formation spread, and engagement axis, while explicit initial states remain available for controlled experiments. Team sizes are not restricted to symmetric one-vs-one encounters. The environment wrappers pack observations, rewards, and done flags for the selected learning-controlled unit identifiers, supporting one-sided training against baselines or control of both teams for self-play.

\begin{table}[H]
\centering
\caption{Comparison of capabilities in the cited public releases. A cross indicates that we did not find documented or direct support for the capability in the cited public release.}
\label{tab:comparison}
\small
\resizebox{\linewidth}{!}{%
\begin{tabular}{lcccc}
\toprule
Capability & BVR Gym & LAG & B-ACE & \system \\
\midrule
Open source & \checkmark & \checkmark & \checkmark & \checkmark \\
JSBSim aircraft dynamics & \checkmark & \checkmark & $\times$ & \checkmark \\
BVR-class missile engagement & \checkmark & $\times$ & \checkmark & \checkmark \\
Mixed aircraft models in one scenario & $\times$ & $\times$ & $\times$ & \checkmark \\
Built-in maneuver-and-fire interface & maneuver only & \checkmark & \checkmark & \checkmark \\
Interchangeable Python/native C++ backends & $\times$ & $\times$ & $\times$ & \checkmark \\
Uniform fixed-width entity table & $\times$ & $\times$ & $\times$ & \checkmark \\
Rule baseline and self-play & \checkmark & \checkmark & \checkmark & \checkmark \\
ACMI/Tacview export or telemetry & \checkmark & \checkmark & $\times$ & \checkmark \\
Real-time 3D visualization & FlightGear & Tacview & Godot & native viewers \\
\bottomrule
\end{tabular}%
}
\end{table}

Table~\ref{tab:comparison} counts a BVR-class engagement only when the released task exposes guided-missile employment at beyond-visual-range separation, rather than merely containing a missile object. LAG includes missile tasks, but its published combat benchmark is WVR-oriented and does not provide a documented BVR workflow. B-ACE uses its own Godot flight model rather than JSBSim, which accounts for the separate cross in the dynamics row.

Mixed aircraft means that distinct flight-dynamics models and their controller parameters can coexist within one episode. Different policies, teams, or initial conditions applied to one common aircraft model do not meet this criterion; the cited public releases document one aircraft model per scenario.

For tactical actions, BVR Gym provides $\operatorname{Box}(3)$ heading, altitude, and throttle commands but fires automatically. LAG uses hierarchical $\operatorname{MultiDiscrete}(3,5,3)$ altitude, heading, and speed commands plus binary fire in missile tasks; B-ACE directly exposes heading, flight level, load factor, and fire. The native \system interface is $\operatorname{MultiDiscrete}(15,15,9,2)$ over heading, altitude, speed, and fire.

The entity-table criterion requires an explicit entity axis with one fixed-width schema. BVR Gym instead returns a $(40,15)$ temporal array; LAG uses flat 15/21-dimensional vectors in 1-vs-1 and 27/33 per agent in 2-vs-2; B-ACE concatenates heterogeneous self/ally/enemy blocks, giving 22 and 41 dimensions. Native \system assigns each aircraft or missile a 40-dimensional row, producing $(3,40)$ and $(5,40)$ in the same scenarios.

Finally, telemetry export and live visualization are separate capabilities. BVR Gym provides ACMI logging and FlightGear visualization, LAG provides Tacview logging, and B-ACE provides an interactive Godot view but no confirmed ACMI/Tacview exporter; \system provides both ACMI output and native real-time viewers.

\section{Learning Interface}
\subsection{High-level action space}
For each controlled aircraft, the default action is
\begin{equation}
 a_t=(k_\psi,k_h,k_v,k_f)\in
 \operatorname{MultiDiscrete}(15,15,9,2),
\end{equation}
where $k_\psi$, $k_h$, and $k_v$ select desired changes in heading, altitude, and speed, and $k_f$ is a binary fire decision. The first three indices are centered and normalized to $[-1,1]$, then scaled to maximum command changes of $45^\circ$, $80$~m, and $80$~m/s, respectively. The resulting command is
\begin{equation}
 u_t=(\Delta\psi_t,\Delta h_t,\Delta v_t),
\end{equation}
which is consumed by the model-specific flight controller. The default interface intentionally defines a simplified tactical abstraction in which target and weapon selection are handled by simulator logic. When $k_f$ requests fire, that logic selects an eligible tracked target and available weapon subject to track and pylon state. Lower-level simulator commands retain explicit target and weapon fields and can support future benchmarks on learned target allocation and weapon assignment.

This design changes what the policy must learn. Direct actuator control exposes aileron, elevator, rudder, and throttle and makes tactical learning depend on a stable low-level controller. In \system, aircraft actuation remains available internally, but the default RL problem begins at the tactical-command layer. Policies can therefore explore formation, track-dependent launch timing, support, and evasion without first solving basic attitude stabilization. Learned target allocation requires the retained lower-level fields or an extended learning interface.

\subsection{Entity-oriented observation}
The native observation represents the tactical scene as a fixed-width table of entities. Each entity row has 40 features and encodes type and validity indicators, relative and absolute kinematics, orientation, team relation, threat and weapon state, and other tactical attributes. Rows are allocated for self, aircraft on both teams, and bounded missile slots, with zero padding for absent objects. For $N_e$ allocated entities, the flattened Gymnasium observation has dimension $40N_e$:
\begin{equation}
 o_i = \operatorname{vec}\left([e_i^{\text{self}},e_{i,1}^{\text{enemy}},\ldots,
 e_{i,M}^{\text{missile}}]\right).
\end{equation}
The fixed-width schema preserves object boundaries and semantic roles instead of merging heterogeneous objects into unrelated flat feature blocks, while remaining compatible with a standard \texttt{Box} space. It is intended to support entity encoders, attention-based policies, masking, and future variable-composition MARL studies. Compact and legacy text/vector observations remain available for compatibility. Experimental examples demonstrate additional energy, missile-warning, and threat-level observation plugins.

\subsection{Rewards and termination}
The reward manager composes independently weighted terms:
\begin{equation}
 r_i=\sum_j w_j r_{i,j}.
\end{equation}
Implemented components cover closing and tracking geometry, distance, altitude advantage and safety, missile launch and duplicate-launch penalties, missile outcome, missile evasion, speed, survival, tactical advantage, and terminal win/loss. Setting a weight to zero disables the corresponding computation. A reward-breakdown API exposes individual terms for diagnosis and plotting. The environment also returns the rule baseline's action as an expert signal, supporting behavior cloning, policy distillation, or auxiliary imitation losses.

An episode ends on the simulator's engagement condition, when all learning-controlled aircraft are destroyed, or at a configured horizon. The information dictionary records the winning team or a draw and exports team rankings expected by multi-agent runners.

\subsection{Framework integration}
The base classes follow Gymnasium-style reset and step conventions and export standard observation and action spaces. Included wrappers adapt the environment to HARL~\cite{zhong2024harl} and MARLBenchmark~\cite{marlbenchmark_offpolicy} APIs, including centralized shared observations for centralized-training/decentralized-execution algorithms. A lightweight \texttt{skrl} PPO entry point is included for installation and smoke testing, while the simulator itself remains algorithm agnostic.

\section{Simulation Components}
\subsection{Aircraft dynamics and control}
Aircraft dynamics are integrated by JSBSim using model-specific XML definitions. \system maintains a local Cartesian north-west-up combat frame and maps simulator state to and from JSBSim geodetic and body-frame properties. When the environment interval exceeds the internal integration limit, each aircraft performs multiple JSBSim substeps. The standard flight controller maps tactical heading, altitude, and speed commands to normalized aileron, elevator, rudder, and throttle commands. Separate parameter sets are registered for the supported fighter families.

\subsection{Missiles and stores}
The pylon manager tracks mounted, remaining, and committed weapons. The weapon factory selects among AIM-120 and alternative parameterized missile models. The generic missile dynamics model includes time-varying mass during motor burn, thrust, gravity, atmospheric density, Mach-dependent drag coefficient, Mach-dependent available normal load, and command-rate limits. Its translational dynamics are
\begin{align}
 m\dot{\mathbf v} &= (T-D)\hat{\mathbf v}+mg n_y\hat{\mathbf s}
 +mg n_z\hat{\mathbf u}+m\mathbf g,\\
 D &= \tfrac12\rho(h)V^2 C_x(M)S,
\end{align}
and are integrated with a fourth-order Runge--Kutta method. This retains the major energy effects that determine whether a long-range intercept remains kinematically feasible. Missile variants can use the built-in AIM-120 implementation or data-driven parameter tables through the generic model.

Table~\ref{tab:guidance} gives a representative matched case from an archived module-level study. An AIM-120C5 was launched 16.21~NM from a straight-flying F-16 target under a high-off-boresight geometry. Changing only the guidance law from proportional navigation (PN) to the multi-stage implementation---initial angular-error correction, lofting, an adaptive navigation coefficient, and terminal PN---shortened time to hit by 8~s and retained more terminal energy. The original trajectory-level archive was unavailable, so this result is a qualitative sanity check rather than statistical validation or operational weapon validation.

\begin{table}[H]
\centering
\caption{Matched high-off-boresight missile-guidance case study.}
\label{tab:guidance}
\small
\begin{tabular}{lccc}
\toprule
Guidance & Range (NM) & Time to hit (s) & Terminal Mach \\
\midrule
Proportional navigation & 16.21 & 36 & 1.35 \\
Multi-stage composite & 16.21 & 28 & 1.71 \\
\bottomrule
\end{tabular}
\end{table}

\subsection{Sensing and opponents}
Aircraft maintain detected enemies, friendly partners, locks, and missile-threat state. Radar parameters are selected by aircraft specification where available, making sensing capability part of the heterogeneous platform definition. Scripted opponents include straight-line, random, aggressive, tactical, standoff, and mutual-assured-destruction-style behaviors. These policies provide curriculum opponents, reproducible evaluation references, and expert actions without requiring a separately trained checkpoint.

\section{Engineering for Reproducible Research}
\paragraph{Dual backends.}
The Python backend is intended for inspection, debugging, and rapid modification. The C++ backend moves aircraft, missile, sensing, baseline, observation, and reward computations into a native core exposed through pybind11. Both are selected through the same Python-level environment family. The backends are not presented as bitwise-identical numerical implementations; rather, they offer a common task abstraction at different points in the transparency--throughput trade-off.

\paragraph{Configuration and extension.}
Complete engagements are described in JSON or JSONC. Researchers can vary force composition, stores, initial geometry, observation type, opponent behavior, episode length, and reward weights without editing the training loop. Observation and reward interfaces are componentized, and repository examples show custom plugins. This makes ablations explicit in configuration and easier to reproduce.

\paragraph{Learning-framework adapters.}
Three adapter paths are released. The MARLBenchmark adapter converts Gymnasium spaces to the legacy Gym API, exposes per-agent spaces, and constructs concatenated shared observations for centralized critics. The HARL adapter returns local and shared observations, rewards, done flags, per-agent information, and an available-action placeholder in the framework's expected tuple; it is exercised end to end by the HAPPO and MAPPO traces in Section~\ref{sec:marl}. A third adapter implements the \texttt{ScenarioConfig} and \texttt{make\_env} contracts used by UHRL, our private branch derived from the public HMP2G framework associated with Unreal-MAP~\cite{hu2026unrealmap}. All three retain the native multidiscrete tactical action and can select the Python or C++ simulator backend. MARLBenchmark and UHRL support here denotes implemented interfaces rather than validation of every algorithm provided by those frameworks.

\begin{figure}[H]
\centering
\includegraphics[width=\linewidth]{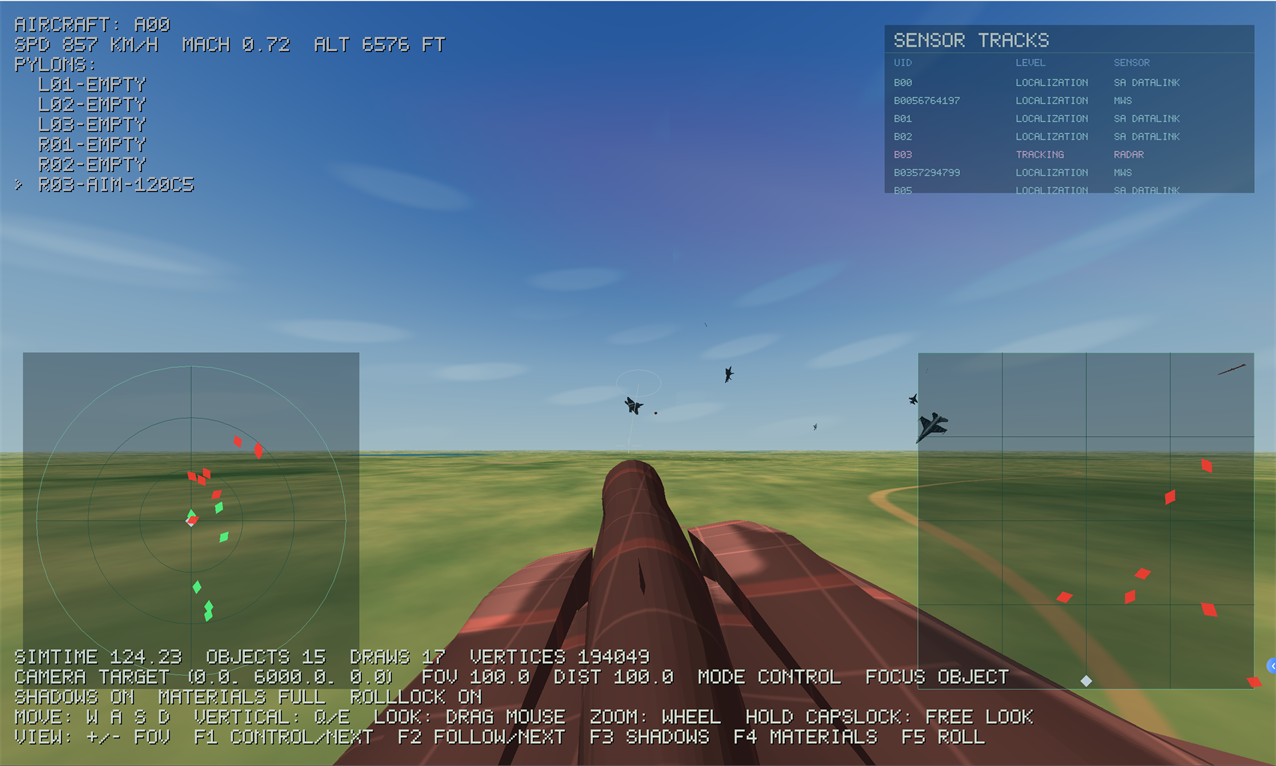}
\caption{Native visualization of a running \system engagement. Rendering is optional and decoupled from headless simulation.}
\label{fig:viewer}
\end{figure}

\paragraph{Telemetry and visualization.}
The simulator can export Tacview-compatible ACMI files and includes Web, OpenGL, and Windows DX11 visualization paths. An experimental UnrealCV bridge mirrors simulator transforms into Unreal Engine. These consumers are decoupled from the headless learning loop, so high-throughput training does not require graphics. Figure~\ref{fig:viewer} shows the native game-mode visualization of a heterogeneous engagement.

\section{Performance Evaluation}
\subsection{Protocol}
We measure single-process environment throughput on Windows 11 using an AMD Ryzen 9 5950X (16 cores, 32 hardware threads). Both backends run headless. Each scenario uses $\Delta t=0.4$~s, tactical scripted opponents, F-16 aircraft carrying six AIM-120C7 missiles each, each backend's entity-observation implementation, reward computation, and team sizes from 1-vs-1 to 10-vs-10. Thus the scenario, aircraft, stores, opponent, and transition workload are matched, while backend-specific numerical and observation implementations remain part of the measured systems. After 20 warm-up transitions, each reported point aggregates three repeats of 500 timed transitions. If an engagement terminates during a repeat, the environment is reset and timing continues; construction, warm-up, and reset time are excluded. ``Steps/s'' denotes complete environment transitions per wall-clock second, not per-aircraft physics updates. The benchmark script and raw per-repeat measurements are distributed with the repository.

\input{benchmark_results.tex}

The principal purpose of this experiment is to characterize scaling with engagement size and quantify the benefit of native execution. The C++ backend is faster than the Python backend in every tested configuration, with mean speedups ranging from $2.72\times$ to $6.55\times$; the largest-scale measurements exhibit substantial run-to-run variance. Performance naturally depends on engagement events: missile launches increase the number of simulated objects and sensing interactions. The 10-vs-10 C++ result is $55.43 \pm 38.71$ steps/s, so its speedup should not be interpreted as a stable asymptotic factor. The benchmark nevertheless shows a consistent native-backend advantage while exercising complete tactical transitions rather than an empty straight-flight loop.

\section{Multi-Agent Learning Integration}
\label{sec:marl}
We use archived HARL runs of MAPPO and HAPPO to verify that \system supports standard end-to-end MARL rather than only multi-aircraft simulation. Both runs use the same \texttt{MultipleCombat-2v2/ShootMissile} task and train two aircraft under centralized training with decentralized execution for approximately 22.5 million environment steps. The plotted metric is the recorded mean episode reward, smoothed with a trailing window of approximately one million environment steps.

\begin{figure}[H]
\centering
\includegraphics[width=0.78\linewidth]{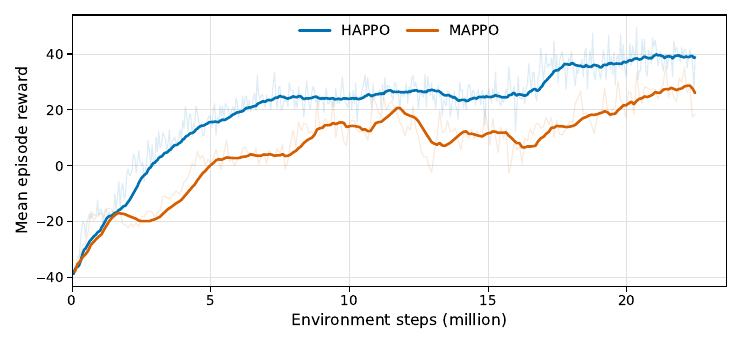}
\caption{Archived 2-vs-2 pure-MARL training traces. HAPPO (seed 3) and MAPPO (seed 1) use the same BVR task. Solid curves are mean episode reward smoothed over approximately one million environment steps; faint curves show the corresponding unsmoothed records, not cross-seed uncertainty. The traces are an integration check, not an algorithm comparison.}
\label{fig:marl-training}
\end{figure}

Both traces show sustained reward improvement over approximately 22.5 million environment steps, supporting compatibility with distinct on-policy MARL implementations and long-horizon multi-agent optimization. Because the runs use different single seeds, their numerical difference is not interpreted. The figure is an integration check rather than evidence of relative performance, convergence, or learned coordination. The extracted records, source hashes, and plotting script are included in the artifact; benchmark-quality comparison still requires multiple seeds and frozen-policy evaluation.

\section{Cross-Aircraft Policy Transfer}
\subsection{Protocol}
We use an archived PPO controller-transfer experiment to test whether the tactical interface preserves its meaning across aircraft. The task is 1-vs-1 against a fixed rule opponent. \emph{PPO-F16} is trained only with the F-16; \emph{PPO-All} is trained with episodes drawn from F-16, F-15, F/A-18, F-22, and F-4N models. At evaluation, each frozen policy controls each of the five aircraft for 960 episodes. The adapted conditions select the registered controller parameters for the active aircraft. In the ``w/o'' controls, every aircraft instead receives the default F-16 controller parameters. Because only aggregate legacy results were recoverable, this experiment supports an environment-level controller claim rather than a multi-seed algorithm comparison; the transcribed values and provenance are included with the paper artifact.

\subsection{Results}
\begin{table}[H]
\centering
\caption{Cross-aircraft 1-vs-1 win rate (\%), 960 episodes per cell. PPO-F16 sees only F-16 during training; PPO-All trains over all five models.}
\label{tab:transfer}
\small
\begin{tabular}{lrrrrrr}
\toprule
Method & F-16 & F-15 & F/A-18 & F-22 & F-4N & Mean \\
\midrule
PPO-F16 & 52.81 & 47.40 & 22.00 & 80.11 & 32.47 & 46.96 \\
PPO-F16 w/o adaptation & 53.13 & 24.90 & 1.88 & 0.00 & 26.04 & 21.19 \\
PPO-All & 65.85 & 52.92 & 33.54 & 77.08 & 47.08 & 55.29 \\
PPO-All w/o adaptation & 65.63 & 35.71 & 0.78 & 0.34 & 44.68 & 29.43 \\
Rule A & 31.88 & 3.82 & 11.41 & 44.31 & 8.75 & 20.03 \\
\bottomrule
\end{tabular}
\end{table}

PPO-F16 transfers without retraining to the four unseen aircraft at a mean win rate of $45.50\%$, compared with $13.21\%$ without aircraft-specific controller adaptation. The largest changes occur on F/A-18 ($22.00\%$ versus $1.88\%$) and F-22 ($80.11\%$ versus $0.00\%$), whose response differs strongly from the default F-16 parameterization. Training over all five models raises the overall mean from $46.96\%$ to $55.29\%$ with adaptation. These results do not establish a universal aircraft-invariant policy or a generally superior PPO method; they show that the same frozen high-level policy can retain useful tactical behavior across multiple JSBSim aircraft models when paired with the corresponding model-specific inner-loop controller.

The maintained lightweight PPO helper now models the simulator's \texttt{MultiDiscrete([15,15,9,2])} action with four categorical branches, rather than optimizing a continuous surrogate that is rounded after sampling. It is distributed as an executable integration baseline. Framework-specific tuning and standardized multi-method comparisons are left to future benchmark work.

\section{Discussion and Limitations}
\system targets research-oriented tactical learning rather than certification-grade or operational combat modeling. In simulation fidelity, aircraft dynamics inherit the limitations of the included JSBSim models; public radar and missile parameters are simplified, and electronic warfare, communications, terrain masking, weather, and detailed countermeasures are incomplete. Native and Python backends share a task abstraction but may differ in implemented models and floating-point trajectories. In learning scope, the default tactical interface delegates target and weapon selection to simulator logic, so learned allocation requires its retained lower-level fields or an extension. In empirical scope, the MARL traces use one run per algorithm without frozen evaluation, controller-transfer evidence is aggregate-only, and the missile result is one module-level case. None supports an algorithmic ranking. Multi-seed frozen MARL evaluation, self-play, explicit coordination analysis, and heterogeneous 5-vs-5 learning remain future work.

These limitations also suggest useful research directions. The per-unit specification can be extended to unmanned combat aircraft, tankers, airborne early-warning platforms, and surface threats. The entity observation naturally supports transformer policies and variable-size masking. Baseline actions and ACMI trajectories provide a path to offline RL and imitation learning. The explicit separation between a headless core and visualization permits richer digital-twin front ends without coupling graphics performance to learning throughput.

\section{Code and Data Availability}
The simulator, configurations, and documentation are available at \url{https://github.com/lizi-Margin/bvr_sim} under the GPLv3 license. The exact code release corresponding to this preprint is the annotated Git tag \texttt{arxiv-v1}. The arXiv supplementary artifact contains the exact paper-side scripts, configurations, aggregate evidence, and a SHA-256 manifest. Full source identifiers and scenario metadata accompany the raw benchmark samples. The artifact also contains the controller-transfer and missile-guidance tables with provenance notes, plus the extracted HARL training records and source hashes used in Figure~\ref{fig:marl-training}. The two legacy physics studies are explicitly identified because their per-episode or trajectory records were not recovered.

\section{Conclusion}
We presented \system as an open environment centered on heterogeneous platform modeling, a shared tactical interface with aircraft-specific controllers, and a practical high-throughput workflow. The native backend remains faster than real time through 10-vs-10, and the HAPPO and MAPPO traces establish compatibility with standard CTDE learners. Cross-aircraft results show that frozen high-level commands retain useful tactical semantics across multiple aircraft when paired with model-specific controllers. These results validate an environment and workflow, not a superior learning algorithm or operational combat model. The released configuration, structured observations, baselines, telemetry, and reproducible artifacts provide a foundation for frozen multi-seed evaluation, explicit coordination studies, and self-play.

\bibliographystyle{plain}
\small
\bibliography{references}

\end{document}

%% file: benchmark_results.tex
\begin{table}[H]
\centering
\caption{Headless throughput (mean $\pm$ standard deviation over three repeats). RTF is simulated time divided by wall-clock time at $\Delta t=0.4$~s.}
\label{tab:throughput}
\begin{tabular}{lrrrr}
\toprule
Scale & Python steps/s & C++ steps/s & C++ RTF & Speedup \\
\midrule
1v1   & $95.84 \pm 8.07$ & $260.65 \pm 2.75$  & 104.26 & $2.72\times$ \\
2v2   & $51.01 \pm 8.09$ & $143.58 \pm 4.56$  & 57.43 & $2.81\times$ \\
4v4   & $22.43 \pm 3.30$ & $88.36 \pm 16.56$  & 35.34 & $3.94\times$ \\
6v6   & $13.07 \pm 1.18$ & $51.21 \pm 12.33$  & 20.48 & $3.92\times$ \\
8v8   & $9.63 \pm 1.24$  & $42.01 \pm 12.06$  & 16.80 & $4.36\times$ \\
10v10 & $8.46 \pm 1.61$  & $55.43 \pm 38.71$  & 22.17 & $6.55\times$ \\
\bottomrule
\end{tabular}
\end{table}